\documentclass[10pt,twocolumn,floats,english,superscriptaddress,nofootinbib]{revtex4-1}
\usepackage{amsthm}
\usepackage{amssymb}
\usepackage{amsmath}
\usepackage{epsf}
\usepackage{dsfont}
\usepackage[T1]{fontenc}
\usepackage[latin9]{inputenc}
\usepackage{lmodern}
\usepackage{geometry}
\usepackage{caption,subcaption}
\usepackage{amsmath}
\usepackage{amssymb}
\usepackage{graphicx}
\usepackage[usenames,dvipsnames]{color}
\usepackage{tikz}
\usepackage{pgfplots}
\usepackage{multirow}
\usepackage{placeins}

\newcommand{\boldx}{{\bf x}}
\newcommand{\boldy}{{\bf y}}
\newcommand{\bolda}{{\bf a}}

\newcommand{\boldzero}{\boldsymbol{0}}

\newcommand{\Z}{\mathbb{Z}}
\newcommand{\R}{\mathbb{R}}
\newcommand{\C}{\mathbb{C}}
\newcommand{\V}{\mathcal{V}}
\newcommand{\sgn}{\operatorname{sgn}}

\theoremstyle{remark}
\newtheorem{remark}{Remark}

\makeatletter
\@ifundefined{textcolor}{}
{%
 \definecolor{BLACK}{gray}{0}
 \definecolor{WHITE}{gray}{1}
 \definecolor{RED}{rgb}{1,0,0}
 \definecolor{GREEN}{rgb}{0,1,0}
 \definecolor{BLUE}{rgb}{0,0,1}
 \definecolor{CYAN}{cmyk}{1,0,0,0}
 \definecolor{MAGENTA}{cmyk}{0,1,0,0}
 \definecolor{YELLOW}{cmyk}{0,0,1,0}
 }

\makeatother

\usepackage{babel}

\begin{document}

\title{Newton Homotopies for Sampling Stationary Points of Potential Energy Landscapes}

\author{Dhagash Mehta}
\email{dmehta@nd.edu}
\affiliation{Department of Applied and Computational Mathematics and Statistics, University of Notre Dame, 
Notre Dame, IN 46556, USA.}
\affiliation{University Chemical Laboratory, The University of Cambridge, Cambridge CB2 1EW, UK.}

\author{Tianran Chen}
\email{chentia1@msu.edu}
\affiliation{Department of Mathematics, Michigan State University, East Lansing, MI 48823, USA.}

\author{Jonathan D. Hauenstein}
\email{hauenstein@nd.edu}
\affiliation{Department of Applied and Computational Mathematics and Statistics, 
University of Notre Dame, Notre Dame, IN 46556, USA.}

\author{David J. Wales}
\email{dw34@cam.ac.uk}
\affiliation{University Chemical Laboratory, The University of Cambridge, Cambridge CB2 1EW, UK.}

\begin{abstract}
\noindent
One of the most challenging and frequently arising problems in many areas of science 
is to find solutions of a system of multivariate nonlinear equations. 
There are several numerical methods that can find many (or all if the system is small enough) 
solutions but they each exhibit characteristic problems. 
Moreover, traditional methods can break down if the system contains singular solutions. 
Here, we propose an efficient implementation of Newton homotopies, 
which can sample a large number of the stationary points of complicated many-body potentials. 
We demonstrate how the procedure works by applying it to the nearest-neighbor
$\phi^4$ model and atomic clusters.
\end{abstract} 

\maketitle

\noindent \textbf{Introduction:} 
Solving nonlinear equations is one of the most frequently arising problems 
in physics, chemistry, mathematical biology and many areas of engineering.
In particular, finding stationary points (SPs) of a potential energy function $V({\bf{x}})$
provides the foundations for global optimization \cite{lis87,walesd97a,waless99}, 
thermodynamic sampling to overcome broken ergodicity 
\cite{BogdanWC06,SharapovMM07,SharapovM07,Wales2013}, as well as rare event dynamics 
\cite{Wales02,Wales04,Wales06,BoulougourisT07,XuH08,TsalikisLBT10,LempesisTBTCS11,TerrellWCH12} 
within the general framework of potential energy landscape theory \cite{Wales03}. 
Here, the SPs of a real-valued function $V({\bf{x}})$
from $\mathbb{R}^{n}$ to $\mathbb{R}$ 
are defined as the simultaneous solutions of the system of equations 
$f_{i}(\textbf{x})=\partial V(\textbf{x})/\partial x_{i} = 0$, for all $i=1,\dots, n$. 
The SPs can be employed to analyze many different properties of a diverse range of physical, 
chemical and biological systems, such as 
metallic clusters, biomolecules, structural glass formers,
and coarse-grained models of soft and condensed matter \cite{Wales03,RevModPhys.80.167}.

Since nonlinear equations are generally difficult to solve, 
it is usually not possible to find all the SPs analytically
and one must resort to numerical methods.
For example, in the Newton-Raphson (NR) approach one refines an initial guess 
via successive iterations in the hope of converging to a solution.
Unfortunately, unless the initial guess is sufficiently close to a solution,
the NR method may converge slowly or diverge.
Furthermore the NR method is also notorious for its erratic behavior 
near singular solutions, e.g., see \cite{griewank1983analysis}.

An alternative method to find SPs is the gradient-square minimization method 
which solves $f_{i}({\bf x})=0$ by minimizing the sum of squares 
$W = \sum_{i=1}^{N} f_{i}(\bf{x})^2$ using traditional numerical methods, 
such as conjugate gradient \cite{angelani2000saddles,broderix2000energy}.
While the minima with $W=0$ are the desired SPs, 
however, the number of minima with $W>0$, 
which are not the solutions of $f_{i}({\bf{x}})$, 
generally outweighs the actual SPs, 
and these non-solutions also have an additional zero Hessian eigenvalue,
making the minimization problem ill-conditioned \cite{DoyeW02,doye2003comment},
and the approach very inefficient in practice \cite{DoyeW02,WalesD03}. 

A systematic approach was proposed in Refs.~\cite{DoyeW02,WalesD03} 
based on eigenvector-following, as implemented in the {\tt OPTIM} package.
This program includes many other geometry optimization techniques,
such as a modified version of the limited-memory Broyden--Fletcher--Goldfarb--Shanno (LBFGS) algorithm \cite{Nocedal80,lbfgs},
single- and double-ended \cite{TrygubenkoW04} transition state searches via 
a variety of gradient-only and second derivative-based eigenvector-following techniques \cite{Wales92,Wales93d},
and hybrid eigenvector-following methods \cite{munrow99,henkelmanj99,kumedamw01}.
The recently described biased gradient squared descent framework 
\cite{:/content/aip/journal/jcp/140/19/10.1063/1.4875477}
may provide a useful alternative, which merits investigation in future work.

Recently, a completely different approach based on algebraic geometry, 
namely the numerical polynomial homotopy continuation (NPHC) method,
has been used to find all the solutions of various models with 
polynomial-like nonlinearity \cite{Mehta:2009,Mehta:2011xs,Mehta:2011wj,Mehta:2011xs,Mehta:2011wj,Kastner:2011zz,Maniatis:2012ex,Mehta:2012wk,Hughes:2012hg,Mehta:2012qr,MartinezPedrera:2012rs,
He:2013yk,Mehta:2013fza,Greene:2013ida}.
After computing an upper bound on the number of 
isolated complex solutions of the given system of equations, 
the system is continuously deformed into a different system whose solution count
agrees with the upper bound.
Then, each solution of the new system is tracked towards
the original system via a single parameter.
This method can identify \textit{all} isolated complex solutions 
(which include real solutions) of the original system (see 
e.g., Refs.~\cite{BertiniBook,Mehta:2011xs,Mehta:2011wj} for more details).
When the number of complex solutions is very large compared to the number of real solutions,
computing all of the real solutions using the NPHC method can be a computationally expensive task.

Another approach to find all the solutions of a system of nonlinear equations 
is an interval based method \cite{gwaltney2008interval}, 
but it has only proved successful for a
very small systems and SPs so far, since it is based on bisections of the ranges.

In this contribution, we use an efficient, robust, and highly parallel implementation
of \emph{Newton homotopies} (NH), a previously underutilized approach for finding SPs.
Unlike the NPHC and the minimization based methods,
the NH approach has the benefit of directly targeting the real SPs.
When compared to the NR method, 
our approach for NH is more effective at finding singular 
solutions and also capable of finding multiple solutions 
starting from a single point.
Numerical experiments with 
nearest-neighbor 2D~$\phi^4$ models and atomic clusters 
suggest that NH is an efficient and 
effective method capable of finding 
a large number of SPs, especially those SPs of higher indices,
within a reasonable amount of time,
and has great potential for use in a wide range of other applications.

\newpage

\noindent \textbf{Newton Homotopy:}
The fundamental goal is to find solutions $\boldx = (x_1,\dots,x_n) \in \R^n$
to a target system consisting of~$n$ equations 
${\bf F}(x_1,\dots,x_n)={\bf F}(\boldx)=\boldzero$.
The general idea of homotopy continuation is to deform the 
\emph{target system} into a different one, 
the \emph{starting system}, for which solutions are easier to compute.
In this article, we focus on deforming using a Newton Homotopy 
developed \mbox{in~\cite{BraninNH,SmaleNH,keller1977global}}
which is given by ${\bf H} : \R^{n+1} \to \R^n$ with
\begin{equation}
    {\bf H}(\boldx,t) := {\bf F}(\boldx) - t {\bf F}(\bolda)
    \label{equ:newton-homotopy}
\end{equation}
for some chosen $\bolda \in \R^n$.
It represents a continuous deformation between the target system ${\bf H}(\boldx,0) \equiv {\bf F}(\boldx)$
and the starting system ${\bf H}(\boldx,1) \equiv {\bf F}(\boldx) - {\bf F}(\bolda)$.
The system of equations ${\bf H}(\boldx,t) = \boldzero$ form a family of solutions
parameterized by $t$ containing the target system ${\bf F}(\boldx)=\boldzero$, which we aim to solve.
  
The $n$ equations ${\bf H}(\boldx,t) = \boldzero$ in
$n+1$ unknowns define the real solution set
\mbox{$\V({\bf H}) := \{ (\boldx,t) \in \R^{n+1} : {\bf H}(\boldx,t) = \boldzero \}$}
containing the target solution set of ${\bf F}(\boldx) = \boldzero$ as a cross-section at $t=0$.
Certifiable methods for numerically tracking along curves in $\V({\bf H})$ are
provided in \cite{Cert1,Cert2}.  

\begin{figure}[ht]
    \centering
    \includegraphics{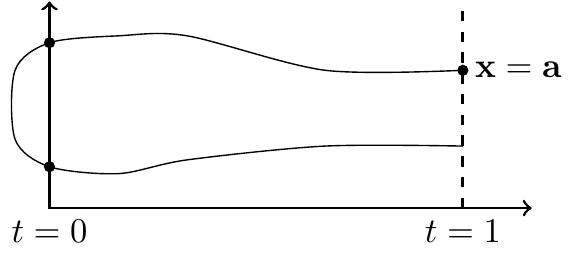}
    \caption{
        A smooth curve defined by ${\bf H}(\boldx,t)=\boldzero$
    }
    \label{fig:newton-homotopy-typical}
\end{figure}

If the Jacobian matrix ${\bf J}_{\bf H}$ of ${\bf H}$ at $(\bolda,1)$ has rank~$n$,
then there is a curve in $\V(\bf H)$ passing through $(\bolda,1)$
that is smooth locally so that one may track along it.
To simplify the situation, assume that ${\bf H}$ \eqref{equ:newton-homotopy}
satisfies the smoothness assumption, namely ${\bf J}_{\bf H}(\boldx,t)$
has rank~$n$ for all $(\boldx,t) \in \V({\bf H})$.
Thus, $\V({\bf H})$ is the union of disjoint smooth curves in $\R^{n+1}$
with one passing through $(\bolda,1)$.
By tracking along this curve, one may locate points in
$\{ (\boldx,t) \in \V({\bf H}) \,:\, t=0 \}$
corresponding to the real solutions of ${\bf F}(\boldx) = \boldzero$.  
Figure \ref{fig:newton-homotopy-typical} depicts~this~situation.

This setup suggests a practical numerical method 
for locating solutions of the system ${\bf F}(\boldx)=\boldzero$:
starting at $(\bolda,1)$, trace the curve defined by ${\bf H}(\boldx,t)=\boldzero$ 
in $\R^{n+1}$ via efficient and reliable numerical methods.
A solution to the target system ${\bf F} = \boldzero$ is produced 
each time the curve passes through the hyperplane at $t=0$.
Since we will not test if the smoothness assumption holds,
we will simply trace along the curve until singularities arise.
Remark~\ref{rmk:singularity} discusses some options
for tracking through singularities.

\noindent \textbf{Tracing smooth curves:}
By the smoothness assumption, the zero set $\V({\bf H})$ of ${\bf H}$ consists of smooth curves.
Let $\gamma$ be the unique curve containing $(\bolda,1)$.
The numerical NH method revolves around 
the procedure of ``tracing'' the curve $\gamma$ from $(\bolda,1)$.
We briefly outline a basic method and refer to standard references,
e.g., \cite{BertiniBook,SW05,AllgowerBook,MorganBook}, for~variations.

For convenience, let $\boldy = (\boldx,t)$ and write ${\bf H}(\boldx,t) = {\bf H}(\boldy)$.
The smooth curve $\gamma$ is naturally parametrized by arc length.
That is, there exists a smooth function $\boldy : \R^+ \to \gamma$ such that $\boldy(0) = (\bolda,1)$,
${\bf H}(\boldy(s)) = \boldzero$, and $\|\dot{{\bf y}}(s)\|_2 = 1$ for all $s \in \R^+$
where $\dot{{\bf y}}(s)$ is the tangent (velocity) vector of the parametrized curve ${\bf y}$ at $s$
and $\|\dot{{\bf y}}(s)\|_2$ is length of this vector.
It represents a trajectory that passes through $(\bolda,1)$, 
satisfies the equation ${\bf H} = \boldzero$, and has unit velocity at all time.
Clearly, parametrizations satisfying these conditions are not unique:
there are at least two going in opposite directions.
Therefore, to trace along a curve without backtracking,
one must be able to determine and maintain a consistent orientation.
It can be shown that under the smoothness condition, the $(n+1) \times (n+1)$ square matrix
$\left[ \begin{smallmatrix} {\bf J}_{\bf H}({\bf y}(s)) \\ \dot{{\bf y}}(s) \end{smallmatrix} \right]$
is never singular, that is, its determinant never vanishes and hence maintains a consistent sign.
Consequently this sign determines the \emph{orientation} of the parametrization.
Once an orientation $\sigma_0 = \pm 1$ is chosen, 
one must keep the orientation consistent while tracing the curve to prevent backtracking.
With the orientation constraint, 
the arc-length parametrization for $\gamma$ is characterized by
\begin{equation}
    \begin{aligned}
    	{\bf J}_{\bf H}({\bf y}(s)) \, \dot{{\bf y}}(s) &= \boldzero, \\
    	\sgn \, \det 
    	    \begin{bmatrix} 
    		    {\bf J}_{\bf H}({\bf y}(s)) \\ \dot{{\bf y}}(s), 
    	    \end{bmatrix} &= \sigma_0, \\
    	\| \dot{{\bf y}}(s) \| &= 1, \\
    	{\bf y}(0) &= (\bolda,1).
    \end{aligned}
    \label{equ:curve-ode-orientation}
\end{equation}

Locally, at any fixed $s \in \R^+$ and its corresponding ${\bf y}(s)$,
the tangent vector $\dot{{\boldy}}(s)$ can be computed efficiently via numerical methods.
In particular, the possible choices for $\dot{{\boldy}}(s)$
can be computed via the $QR$-decomposition of the transpose matrix ${\bf J}_{\bf H}(\boldy(s))^T$.
Furthermore, utilizing the information produced during the $QR$-decomposition,
the correct choice of $\dot{{\boldy}}(s)$ can be made, as a by-product,
with at most $O(n)$ extra floating point operations.

Globally, in principle, any ordinary differential equation solver capable of integrating the above system 
can be used to trace the curve and potentially obtain solutions to the target system point at $t=1$.
Numerical methods based on this idea are generally referred to as ``global Newton methods''~\cite{SmaleNH}.
Our implementation employs a ``prediction-correction scheme'' due to numerical stability concerns
\cite{keller1977global}.

\begin{remark}
    \label{rmk:singularity}
    The numerical method described above is actually 
    capable of handling cases where the curve $\gamma$ contains isolated singularities,
    such as points at which two curves intersect transversally.
    More advanced techniques for handling singularities can be found 
    in~\cite{AllgowerBook,keller1977numerical,rheinboldt1978numerical,georg1981tracing,glowinski1985continuation,Hao:2012}.
\end{remark}

\begin{figure}[h!]
    \centering
    \includegraphics{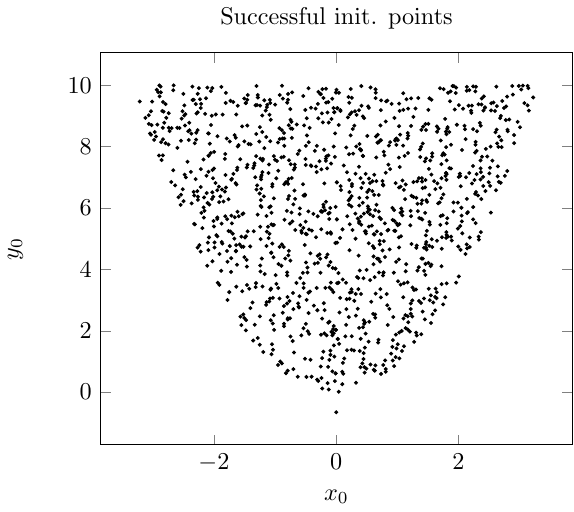}
    \caption{
	    Scatter plot of some starting points $(x_0,y_0)$ for which the NH 
	    \eqref{equ:newton-homotopy} was successful in obtaining the singular solution 
	    $(0,0)$ of the system \eqref{eq:Griewank-Osborne} within machine precision.
    }
    \label{fig:go-scatter}
\end{figure}

\noindent \textbf{An Example System:}
Consider the system
\begin{equation}
    \begin{cases}
	\frac{29}{16}x^3 - 2 x y &= 0, \\
	y - x^2 &= 0.
    \end{cases}
    \label{eq:Griewank-Osborne}
\end{equation}
from \cite{griewank1983analysis}.  
This system has only one solution in $\R^2$, namely $(0,0)$, which has multiplicity $3$.
It is shown in \cite{griewank1983analysis} that starting from
almost every point in $\R^2\setminus\{(0,0)\}$, the NR method will diverge.
In other words, the NR method will almost surely never find the solution of this system.
Figure~\ref{fig:go-scatter} shows that the NH method 
\eqref{equ:newton-homotopy} was successful at locating the solution 
for many starting points $(x_0,y_0)$.

\noindent \textbf{The Nearest-Neighbor Two-dimensional $\phi^4$ Model:} 
We consider a model from theoretical physics called the two-dimensional nearest-neighbor $\phi^4$ model. 
It has been widely studied because it is one of the simplest models with a continuous configuration space 
that exhibits a phase transition in the same universality class as the two-dimensional Ising model. 
For an $N \in \Z^+$ and $J,\lambda,\mu \in \R$ the model, 
in $N^2$ variables $\boldx = (x_{11}, x_{12}, \dots , x_{NN})$, is $V(\boldx)$ given by 
\begin{equation}
\mbox{\footnotesize $
V(\textbf{x}) = \displaystyle\sum_{(i,j) \in \Lambda}\left(
	\frac{\lambda}{4!}x^{4}_{ij} - 
	\frac{\mu^2}{2}x^{2}_{ij} + 
	\frac{J}{4}\sum_{(k,l) \in \mathcal{N}_{(i,j)}}(x_{ij}-x_{kl})^2
    \right)$}\label{equ:phi4}
\end{equation}
where $\Lambda \subset \Z^{2}$ is the standard square lattice with $N^2$ 
lattice sites and $\mathcal{N}_{(i,j)} \subset \Lambda$ 
denotes the four nearest-neighbor sites of $(i,j)$. 
The $N^2$ stationary equations are given by
\begin{equation}
    \frac{\partial V(\boldx)}{\partial x_{ij}} = 
    \frac{\lambda}{3 !} x^3_{ij} + (4J - \mu^2) x_{ij} - \sum_{(k,l) \in \mathcal{N}_{(i,j)}} J x_{kl} = 0
    \label{equ:phi4-stationary}
\end{equation}
for each pair of $i,j = 1,\dots,N$.
Given the physical context, only real solutions are needed. We use periodic boundary conditions, $\lambda = 3/5$ and $\mu^2 = 2$.

A variety of computational tools have been used to study this model.
In particular, the NPHC method has found all the SPs for $N=3,4$ 
in a previous study \cite{Kastner:2011zz,PhysRevE.85.061103}.
However, this family of problems poses a particularly tough challenge
to methods that find all complex solutions, since
the total number of solutions in $\C^{N^2}$, counting multiplicity,
is always equal to its total degree (the Bezout bound)~$3^{N^2}$,
which grows quickly as $N$ increases.
Direct computation of all complex solutions becomes
unfeasible as $N$ increases.
However, by varying the parameter $J$ from 0 to 1, 
we pass from the case when all the solutions are real 
to where only an extremely small fraction are real.
For the latter limit
the NH approach, which directly targets the real solutions,
has a clear advantage over methods that compute~all~complex~solutions.

\begin{figure}[h!]
    \centering
    \includegraphics{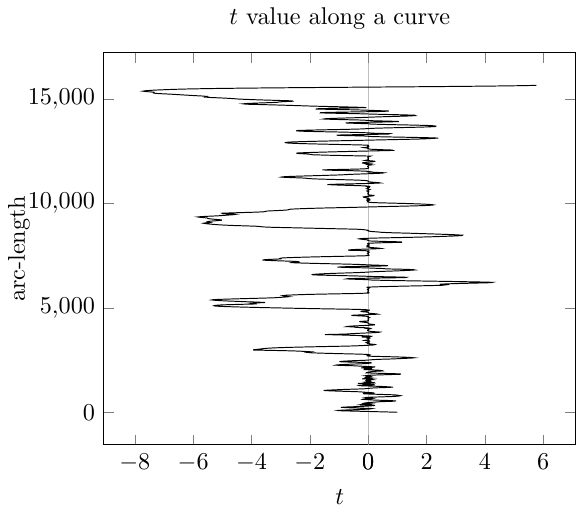}
    \caption{
	    The $t$ value along a curve defined by the Newton homotopy for 
	    \eqref{equ:phi4-stationary} for $N=6$ and $J=0.9$.
	    The vertical axis represents the arc-length,
	    that is, the distance traveled along the curve.
	}
    \label{fig:n6-j09-t}
\end{figure}
In our numerical experiments, Newton homotopies 
\eqref{equ:newton-homotopy}
were applied to \eqref{equ:phi4-stationary} with varying values for $N$ and $J$.
From a \emph{single} randomly chosen starting point
multiple real solutions were obtained.
Table~\ref{tab:phi4-single} summarizes the ability and efficiency of 
NH in finding the real solutions for a range of $N$ and $J$ values.
Indeed, \emph{all real solutions were found in many cases}.
For example, with $N=3$, in the case of $J=0.9,\ 0.8,\ 0.7,\ 0.6,\ 0.5,$ and $0.4$,
our NH implementation was able to obtain \emph{all} of them 
with a single randomly chosen starting point.
The CPU time information in the table corresponds to
a workstation with a 3.4\,GHz \textsf{Intel Core i5-3570K} processor.
The results highlight the strength of the NH:
it is capable of finding a large number of real solutions very quickly.
The efficiency is particularly noteworthy in the case of $N=7$ and $N=8$.
With a total of more than $10^{23}$ and $10^{30}$ complex solutions, respectively,
any approach that aims to find all complex solutions is clearly impractical. 
In contrast, with $J=0.9$, the NH method was able to obtain 358 and 1522 real solutions,
for the cases of $N=7,8$ respectively, using a \emph{single} starting point \emph{within 1 minute}.

\begin{table}[h!]
    \centering
    \scriptsize
    \begin{tabular}{|c|c|r|r|r|}
    	\hline
    	                              $N$                                &  $J$   & No. of SPs & \% of total SPs found &   Time \\ \hline\hline
    	                       \multirow{5}{*}{3}                        & $0.90$ &          3 &           (All) 100\% & 0.008s \\
    	                                                                 & $0.70$ &          3 &           (All) 100\% & 0.012s \\
    	                                                                 & $0.50$ &        171 &           (All) 100\% & 0.999s \\
    	                                                                 & $0.30$ &       1121 &                99.1\% & 2.001s \\
			    
	\hline
	\multirow{4}{*}{4} & $0.90$ &         83 &           (All) 100\% & 0.903s \\
    	                                                                 & $0.60$ &        199 &                68.4\% & 1.371s \\
    	                                                                 & $0.30$ &      40225 &                40.6\% & 59.27s \\ \hline
    	                       \multirow{3}{*}{5}                        & $0.90$ &        102 &                     - & 2.009s \\
    	                                                                 & $0.60$ &        679 &                     - & 49.50s \\ \hline
    	                       \multirow{2}{*}{6}                        & $0.90$ &        208 &                     - & 23.95s \\
    	                                                                 & $0.60$ &        959 &                     - & 52.37s \\ \hline
    	                       \multirow{2}{*}{7}                        & $0.90$ &        358 &                     - & 29.66s \\
    	                                                                 & $0.60$ &       3266 &                     - & 37.25s \\ \hline
    	                       \multirow{2}{*}{8}                        & $0.90$ &        674 &                     - & 43.12s \\
    	                                                                 & $0.60$ &       1538 &                     - & 55.99s \\ \hline
    \end{tabular}
    \caption{
	    The number of real solutions of \eqref{equ:phi4-stationary}
	    found using NH with \emph{one} starting point. The percentages are computed
	    with respect to all SPs~\cite{Kastner:2011zz,PhysRevE.85.061103}.
    }
    \label{tab:phi4-single}
\end{table}
These cases also highlight the ability of NH to obtain 
multiple solutions using a single starting point.
Figure \ref{fig:n6-j09-t} illustrates the $t$-value along the single curve defined by the
Newton homotopy for \eqref{equ:phi4-stationary} with $N=6$ and $J=0.9$.
Here, $t$ (horizontal axis) is plotted against the arc-length (vertical axis).
Note the numerous crossings of the hyperplane at $t=0$,
represented by the light vertical line in the middle.
Each crossing produces a distinct real solution for \eqref{equ:phi4-stationary}.

Using multiple starting points and tracing multiple curves,
the likelihood for the NH method to obtain many or all
real solutions can be improved substantially.
Note that the curve tracings are completely 
independent and hence can be performed in parallel.
Table \ref{tab:phi4-many} summarizes the efficiency of the NH method
in finding a large number of real solutions for \eqref{equ:phi4-stationary}
using multiple randomly chosen starting points.
The timing information is based on the performance
on a cluster of 32 nodes, each having a quad-core \textsf{Intel Xeon} 
processor running at 2.4\,GHz.

\begin{table}[h!]
    \centering
    \scriptsize
    \begin{tabular}{|c|c|c|r|r|r|}
	    \hline
	    $N$                 & $J$    & No. of Start Points & No. of SPs & \% SPs & Time \\ \hline \hline
	    \multirow{4}{*}{4}  & $0.90$ & 1000                & 83                       & (All) 100\%   & 7.15s  \\
	                        & $0.60$ & 1000                & 291                      & (All) 100\%   & 110.50s  \\
	                        & $0.30$ & 1000                & 99187                    & (All) 100\%   & 121.01s  \\ \hline
	    \multirow{3}{*}{5}  & $0.90$ &  500                & 243			          & -             & 99.50s  \\
		                    & $0.60$ &  500                & 1083                     & -             & 139.21s  \\
			                & $0.30$ &  500                & 30971                    & -             & 353.97s  \\ \hline
	    \multirow{2}{*}{6}  & $0.90$ &  100                & 579                      & -             & 47.33s  \\
			                & $0.60$ &  100                & 4172                     & -             & 329.15s  \\ \hline
	    \multirow{2}{*}{7}  & $0.90$ &   64                & 917                      & -             & 61.19s  \\
                            & $0.60$ &   64                & 3965                     & -             & 86.31s  \\ \hline
	    \multirow{2}{*}{8}  & $0.90$ &   32                & 1522                     & -             & 58.70s  \\
                            & $0.60$ &   32                & 5694                     & -             & 61.11s  \\ \hline
    \end{tabular}
    \caption{
	    The number and percentage of solutions for \eqref{equ:phi4-stationary}
	    found using Newton homotopy with many starting points.
    }
    \label{tab:phi4-many}
\end{table} 

\FloatBarrier

\noindent \textbf{Lennard-Jones Clusters:}
We now apply the NH method to finding SPs of atomic clusters 
of $N$ atoms bound by the Lennard-Jones potential \cite{jonesi25},
which is defined as
\begin{equation}
    V_N = 4\epsilon \sum_{i=1}^N \sum_{j=i+1}^N 
	\left[ 
	    \left( \frac{\sigma}{r_{ij}} \right)^{12} - 
	    \left( \frac{\sigma}{r_{ij}} \right)^{6} 
	\right],
    \label{equ:lj-cluster}
\end{equation}
where $\epsilon$ is the pair well depth, 
$2^{1/6}\sigma$ is the equilibrium pair separation, and
$r_{ij} = \sqrt{(x_i - x_j)^2 + (y_i - y_j)^2 + (z_i-z_j)^2}$ is the distance
between atoms $i$ and $j$.
We take $\epsilon = \sigma = 1$.
Defined in terms of the distances,
$V_N$ is clearly invariant under rotation and translation.
Therefore we can fix
$x_1 = y_1 = z_1 = y_2 = z_2 = z_3 = 0$.
Hence, there are in total $3N-6$ variables in $V_N$ yielding $3N-6$ stationary equations $\nabla V_N = {\bf 0}$.
For this model, an extensive search for minima and saddle points was carried out
in \cite{DoyeW02} for $N$ up to $13$,
and a search for minima and saddles of index one
(transition states) for $N = 14$ was presented in \cite{2005JChPh.122h4105D}.
Table~\ref{tab:lj-cluster} shows that NH
can find a large number of SPs for \eqref{equ:lj-cluster} at each $N$ value.
It is worth noting that the results suggest
the NH approach is particularly useful in finding SPs of higher Morse indices
(the number of negative eigenvalues of the Hessian matrix of $V_N$):
among the SPs found, the majority have Morse index near the middle
of the possible range (from 0 to the number of variables, $3N-6$), 
which may be attributed to the fact that there are exponentially more 
SPs in the mid-range of the indices than at the extremes (index $0$ and index $3N-6$).

Though the number of SPs shown in this table is much less than
    the known collections of SPs found in \cite{DoyeW02} and \cite{2005JChPh.122h4105D},
    the result is still encouraging since,
    as a demonstration of the effectiveness of the NH approach,
    we have restricted the computation time to only 24 hours in each case.
    Given the parallel nature of this approach,
    the number of SPs that can be obtained will likely be significantly improved 
    when more time and computational resource are used.

\vspace{6pt}

\begin{table}[h]
    \centering
    \scriptsize
    \begin{tabular}{|c|c|c|c|}
    	\hline
    	    &       \#            &\# local    &\# transition \\
    	$N$ & SPs / Energy levels &   minima   &     states      \\ \hline\hline
    	 3  &       9 /   4       &     3      &        1        \\ \hline
    	 4  &      31 /  11       &     3      &        3        \\ \hline
    	 5  &      101 /  39      &     1      &        5        \\ \hline
    	 6  &      204 / 148      &     2      &        6        \\ \hline
    	 7  &      725 / 265      &     4      &       13        \\ \hline
    	 8  &      597 / 224      &     8      &        1        \\ \hline
    	 9  &      991 / 501      &     16     &        1        \\ \hline
    	10  &     2510 / 546      &     22     &       71        \\ \hline
    	11  &     9940 / 552      &     34     &       83        \\ \hline
    	12  &     20994 / 623     &     62     &       90        \\ \hline
    	13  &     10920 / 289     &     73     &       92        \\ \hline
    	14  &     32517 / 264     &     37     &       81        \\ \hline
    \end{tabular}
    \caption{
	    Number of SPs and distinct energy levels of 
	    \eqref{equ:lj-cluster} found using Newton homotopies.
    }
    \label{tab:lj-cluster}
\end{table}
\FloatBarrier

\noindent \textbf{Conclusion:} 
We have developed a novel implementation 
of the Newton homotopy method which, in our experiments,
is much more efficient at finding SPs of PELs arising in 
chemical physics than the usual Newton-Raphson method.
Newton homotopies appear to be better behaved
at possible singular SPs. 
Our results suggest that the NH method has the 
potential to replace the NR method in many 
contemporary computational approaches, especially in computational chemistry.

\noindent \textbf{Acknowledgements:} DM and DJW were supported by the EPSRC and the ERC.
DM and JDH were supported by the DARPA YFA program.
JDH and TC were supported by NSF DMS-1262428 and
DMS-1115587, respectively.


\end{document}